\documentclass[submitting]{nst}

\usepackage{newtxtext,newtxmath}
\usepackage{subfigure,dcolumn}
\usepackage[T2A,T1]{fontenc}
\usepackage[russian,english]{babel}
\usepackage[utf8]{inputenc}
\usepackage{nucleardata}
\usepackage{microtype}
\usepackage{graphicx}
\usepackage{color}
\usepackage{placeins}
\usepackage[normalem]{ulem}
\usepackage[capitalize]{cleveref}
\usepackage{bm}
\usepackage{siunitx}
\usepackage{xcolor}
\usepackage{booktabs}
\usepackage{multirow}

\newcommand{\red}[1]{\textcolor{black}{#1}}

\begin{document}

\title{Constraining cross sections for unstable $^{153,159}$Gd$(n,\gamma)$ and their astrophysical implications}

\author{Shu-Tong Zhang}
\affiliation{School of nuclear Science and Technology, University of South China, Hengyang 421001, China}

\author{Zhi-Cai Li}
\email{lizhicai@usc.edu.cn}
\affiliation{School of nuclear Science and Technology, University of South China, Hengyang 421001, China}
\affiliation{Key Laboratory of Advanced Nuclear Energy Design and Safety, Ministry of Education, Hengyang 421001, China}

\author{Kai-Jun Luo}
\affiliation{School of nuclear Science and Technology, University of South China, Hengyang 421001, China}

\author{Hong-Chen Liu}
\affiliation{School of nuclear Science and Technology, University of South China, Hengyang 421001, China}

\author{Yun-Jie Guo}
\email{413454945@qq.com}
\affiliation{Office of the Financial Commission, CPC Hunan Provincial Committee, Changsha 410011, China}

\author{Kai-Xin Zhao}
\affiliation{School of nuclear Science and Technology, University of South China, Hengyang 421001, China}

\author{Zi-Ang Lin}
\affiliation{School of nuclear Science and Technology, University of South China, Hengyang 421001, China}

\author{Wen Luo}
\email{wenluo-ok@163.com}
\affiliation{School of nuclear Science and Technology, University of South China, Hengyang 421001, China}
\affiliation{Key Laboratory of Advanced Nuclear Energy Design and Safety, Ministry of Education, Hengyang 421001, China}

\begin{abstract}

Neutron capture $(n,\gamma)$ cross sections of Gadolinium (Gd) isotopes are critical to astrophysics research, nuclear reactor designs, and medical applications. However, the available $(n,\gamma)$ data on unstable Gd isotopes are scarce and direct measurement is challenging. 
In this work, we propose an approach to infer the $(n,\gamma)$ cross sections for unstable $^{153,159}$Gd isotopes by constraining both the $\gamma$-ray strength functions ($\gamma$SFs) and nuclear level densities (NLDs). 
Specifically, the key $\gamma$SF parameters are adjusted to match the available experimental data, and the NLD parameters are determined by renormalizing microscopic level densities through a Bayesian optimization method. Our approach is verified by comparing our predictions with the experimental $(n,\gamma)$ data for the stable $^{155,157}$Gd isotopes. We then infer the unstable $^{153,159}\text{Gd}(n,\gamma)$ cross sections within the neutron energy range of 0.01--5.0 MeV. The resulting uncertainty is about $30\%$, which is significantly reduced by a factor of 5.5 compared to a large uncertainty of $\sim167\%$ predicted with different nuclear models in TALYS. We further calculate the astrophysical reaction rates for the $^{153,159}\text{Gd}$ isotopes. It is found that the $^{159}\text{Gd}(n,\gamma)$ rate is larger by a factor of $\sim$2.9 than the JINA REACLIB recommendation. This enhancement increases the neutron capture branching ratio at $^{159}$Gd. Consequently, the resulting $^{160}$Gd abundance is increased by a factor of $\sim$2 compared to predictions using the JINA REACLIB rate in $s$-process nucleosynthesis simulations. Our approach is promising for extracting $(n,\gamma)$ data on a wider range of unstable isotopic chains as well as for essential astrophysical reaction network calculations and nuclear science applications.

\end{abstract}


\maketitle

\nolinenumbers

\section{Introduction}
\label{sec:introduction}

\begin{figure}[h]
  \includegraphics[width=\columnwidth]{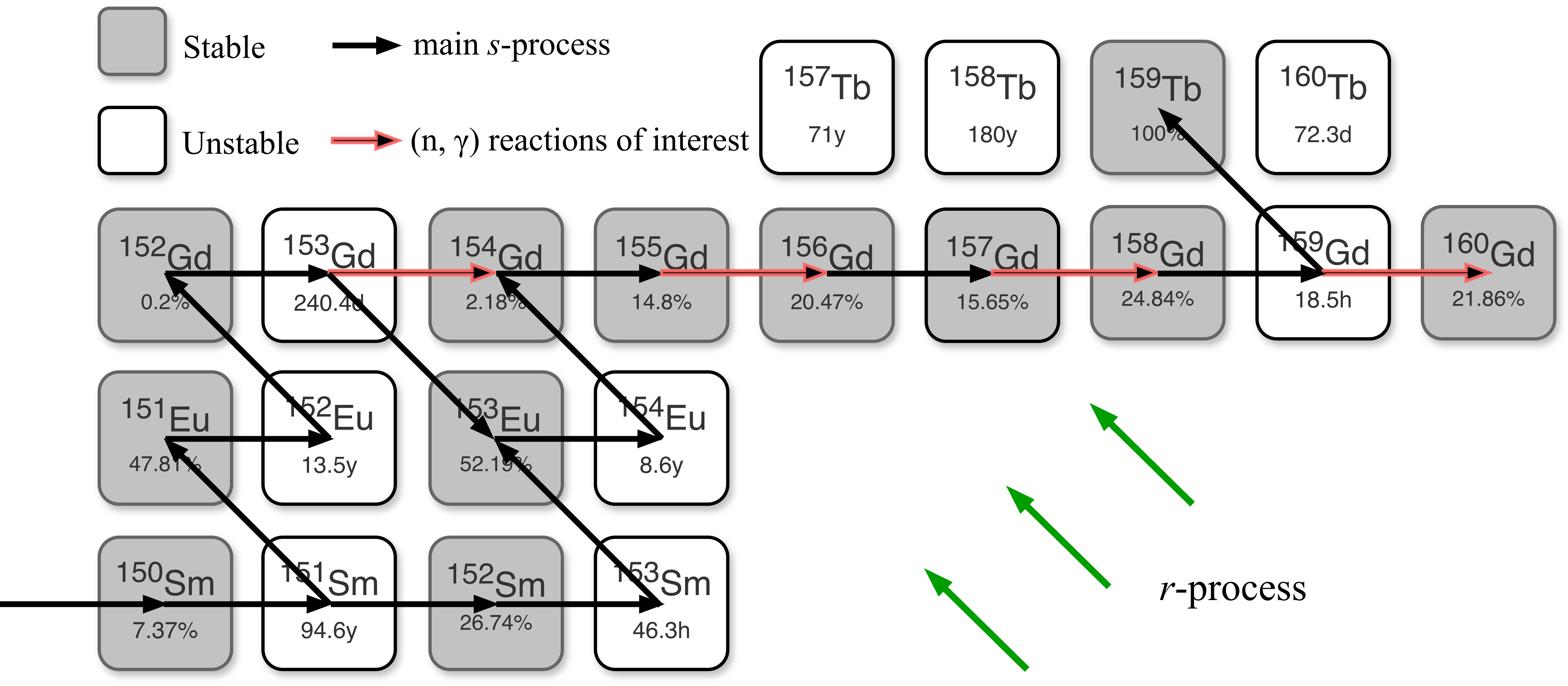}
  \caption{
  Simplified nucleosynthesis path of Sm, Eu, Gd, and Tb during the $s$-process. Stable isotopes are shown in grey, while unstable isotopes are in white. Their natural abundances or half-lives are listed accordingly. The black arrows indicate the main $s$-process flow. The red-framed black arrows indicate the $(n,\gamma)$ reactions of interest investigated in this work.  
  }
  \label{fig:Gd_process}
\end{figure}

Most of the elements heavier than iron are produced by a sequence of neutron capture reactions and $\beta$-decays during different phases of stellar evolution. The two dominant processes involved are the rapid ($r$) and the slow ($s$) neutron capture processes~\cite{mumpowerImpactIndividualNuclear2016,arconesWhitePaperNuclear2017,burbidge1957fowler,mazzoneMeasurement154GdngCross2020}. In particular, unlike most of the isotopes, $^{152}$Gd and $^{154}$Gd receive contributions only from the $s$-process. These so-called ``s-only'' isotopes are crucial for testing $s$-process models. 
As illustrated in Fig.~\ref{fig:Gd_process}, they are shielded against the $\beta$-decay chains from the $r$-process region by stable samarium isobars~\cite{mazzoneMeasurement154GdngCross2020,wisshakStellarNeutronCapture1995,Massimi_2020581}. The determination of their final abundances, however, crucially depends on the branchings at unstable nuclei, where neutron capture rates competes with $\beta$-decay rates~\cite{scielzoDetermininigNgCross2008,wisshakStellarNeutronCapture1995}.
In the standard $s$-process, the branching at unstable isotope $^{153}$Gd is relatively minor~\cite{wisshakStellarNeutronCapture1995}. In contrast, under the higher neutron density conditions of the intermediate neutron capture process ($i$-process), the contribution from $^{153}$Gd$(n,\gamma)$ may become non-negligible. Similarly, a comparable situation exists for the unstable isotope $^{159}$Gd, which thus requires further analysis. Furthermore, $^{155}$Gd and $^{157}$Gd account for the majority of the natural Gd cross section, with $^{157}$Gd notably possessing the highest thermal cross section of 254,000 barns, making them pivotal for reactor designs and medical applications such as Gadolinium Neutron Capture Therapy~\cite{kangNeutronCaptureMeasurements2015,osti_10126890,hoGadoliniumNeutronCapture2022}.
Consequently, obtaining reliable neutron capture cross sections and reaction rates for Gd isotopes is critical for stellar nucleosynthesis models, reactor designs, and medical applications.

While experimental cross section data for stable Gd isotopes are generally available, experimental data for the crucial unstable isotopes $^{153}$Gd and $^{159}$Gd are scarce and direct measurements remain extremely challenging.
Specifically, measurements involving radioactive targets are hindered by the high radioactivity of the samples. Furthermore, experiments in inverse kinematics are currently not feasible for $(n,\gamma)$ reactions because free-neutron targets are not available \cite{ratkiewiczNeutronCaptureExotic2019,thapa2025constrainingcapturecrosssections}. 
In the absence of experimental data, most neutron capture cross sections and reaction rates required for stellar modeling rely heavily on theoretical model predictions. However, these predictions often suffer from large uncertainties due to difficulties in accurately describing the deexcitation process of the nucleus formed after neutron capture. Indeed, this process is ruled by fundamental nuclear properties ($\gamma$-ray strength functions ($\gamma$SF), nuclear level densities (NLD), etc.), for which existing nuclear models yield widely conflicting results in the absence of experimental constraints. As a result, the predicted cross sections can exhibit significant divergence, reaching variations of up to several orders of magnitude~\cite{PhysRevLett.116.242502, ARNOULD200797, perezsanchezSimultaneousDeterminationNeutronInduced2020, zhangConstrainingNeutronCapture2025}. 
Consequently, these theoretical uncertainties propagate directly into astrophysical reaction rates, limiting the reliability of nucleosynthesis simulations for unstable isotopes.

In this work, we present an effective approach to infer the $(n,\gamma)$ cross sections for unstable $^{153,159}$Gd isotopes by constraining both the $\gamma$SFs and NLDs. The constrained $\gamma$SFs and NLDs are implemented in the nuclear reaction code TALYS-2.0~\cite{koningTALYSModelingNuclear2023}. We validate the approach using the stable isotopes $^{155,157}$Gd, for which experimental data are available, and then infer the $(n,\gamma)$ cross sections for the unstable isotopes $^{153,159}$Gd.
\red{Furthermore, we calculate the resulting astrophysical reaction rates for $^{153,159}$Gd and explore the implications for the $s$-process branching at $^{159}$Gd and the production of $^{160}$Gd.}
The remainder of this paper is structured as follows: In Sec.~\ref{sec:extraction}, we detail our approach for constraining the $\gamma$SF and NLD parameters. 
In Sec.~\ref{sec:cs}, \red{We validate our approach using stable $^{155,157}$Gd and present the calculated $(n,\gamma)$ cross sections for the unstable $^{153,159}$Gd isotopes.
In Sec.~\ref{sec:astro}, we discuss the astrophysical implications, including the reaction rates for $^{153,159}$Gd, the $s$-process branching at $^{159}$Gd, and the impact on the production of $^{160}$Gd.} In Sec.~\ref{sec:summary}, we give a conclusion and outlook.

\section{EXTRACTION OF $\gamma$-ray STRENGTH FUNCTION AND NUCLEAR LEVEL DENSITY}
\label{sec:extraction}

\subsection{$E1$ $\gamma$-Ray Strength Function}
\label{sec:E1_SF}

The $\gamma$SF is important for the description of any transition involving gamma rays in nuclear reactions, providing essential information for predicting capture cross sections, $\gamma$-ray production spectra, and competition between $\gamma$-ray and particle emission \cite{capoteRIPLReferenceInput2009}. 
In this section, we extract the $E1$ $\gamma$SF parameters for the $^{154,156,158,160}$Gd isotopes, which serve as the compound nuclei for the neutron capture reactions calculated in this work. These parameters are constrained using experimental $\gamma$SF data from the IAEA Photon Strength Function Database~\cite{gorielyReferenceDatabasePhoton2019}, derived from photo-absorption and photo-neutron cross section measurements \cite{vasilev1971giant,PhysRevC.2.1951,GUREVICH1981257,osti_4716950,GREENWOOD1978327}.
Specifically, we utilize the data in the giant dipole resonance (GDR) region above the neutron separation energy ($S_n$) for extracting.

The Standard Lorentzian (SLO) model \cite{brinkIndividualParticleCollective1957,axelElectricDipoleGroundState1962} is probably the most widely adopted approach for describing photo-absorption data of medium-weight and heavy nuclei \cite{bartholomewGammaRayStrengthFunctions1973,bermanMeasurementsGiantDipole1975,dietrichAtlasPhotoneutronCross1988}. The SLO form for the $\gamma$-ray strength is given by
\begin{equation}
  f_{X\ell}(E_\gamma)
  = K_{X\ell}
    \sum_{i=1}^{N_{\text{peak}}}
    \frac{\sigma_i\,E_\gamma\,\Gamma_i^2}{(E_\gamma^2 - E_i^2)^2 + E_\gamma^2\,\Gamma_i^2},
  \label{eq:psf}
\end{equation}
where $K_{X\ell} = [(2\ell+1)\pi^{2}(\hbar c)^{2}]^{-1}$, $X$ denotes either electric ($E$) or magnetic ($M$) radiation, and $\ell$ is the multipolarity of the transition. $E_i$, $\Gamma_i$, and $\sigma_i$ denote the centroid energy, width, and peak cross section of the $i$th resonance, respectively. According to the Reference Input Parameter Library (RIPL-3, Ref.~\cite{capoteRIPLReferenceInput2009}), nuclei with quadrupole deformation parameter $\beta_2 \le 0.01$ behave as spherical. The Gd isotopes studied here ($\beta_2 \in [0.3,0.35]$ from the Skyrme-Hartree-Fock-Bogoliubov mass model of Goriely and Pearson \cite{gorielySkyrmeHartreeFockBogoliubovNuclearMass2009}) were treated as deformed and therefore described with two Lorentzian components ($N_{\text{peak}} = 2$).

The parameter set is $\theta = (\sigma_1,E_1,\Gamma_1,\sigma_2,E_2,\Gamma_2)$, and it was determined by minimizing the chi-square defined as
\begin{equation}
  \chi^2(\theta) = \sum_k \frac{\left[f_{\text{exp}}(E_k) - f_{E1}(E_k;\theta)\right]^2}{\sigma_k^2},
\end{equation}
where $f_{\text{exp}}(E_k)$ is the experimental $\gamma$SF value at the $k$-th energy point $E_k$, and $\sigma_k$ is the corresponding experimental uncertainty. The minimization was performed using nonlinear least-squares method. 

\begin{figure*}[!ht]
  \includegraphics[width=0.85\textwidth]{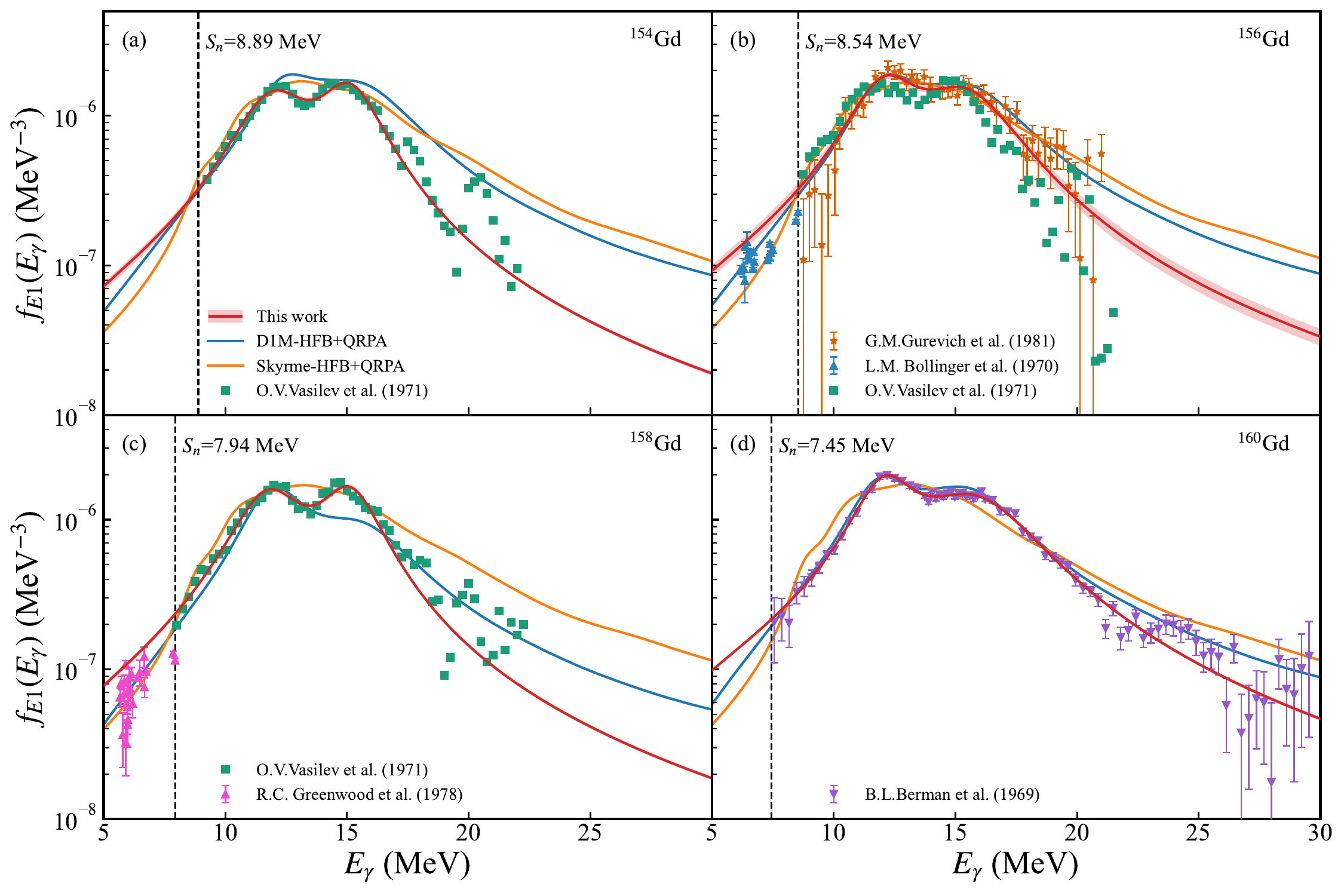}
  \caption{
  Panels (a)--(d) present the $\gamma$SF  for $^{154,156,158,160}$Gd. Our SLO fits are shown as a red line with one standard deviation uncertainty bands. These are compared with the experimental $\gamma$SF from Refs.~\cite{vasilev1971giant,PhysRevC.2.1951,GUREVICH1981257,osti_4716950,GREENWOOD1978327} and microscopic model evaluations (D1M-HFB+QRPA\cite{gorielyGognyHFB+QRPADipoleStrength2018} and Skyrme-HFB+QRPA\cite{gorielyMicroscopicHFBQRPA2004}, depicted by blue and orange lines, respectively). The vertical dashed lines indicate the $S_n$ for each isotope.
  }
  \label{fig:rsf_fit}
\end{figure*}

\cref{fig:rsf_fit} shows our $E1$ $\gamma$SF fits for $^{154,156,158,160}$Gd compared with two microscopic predictions: the Skyrme-HFB+QRPA strength \cite{gorielyMicroscopicHFBQRPA2004} and the D1M-Gogny HFB+QRPA strength \cite{gorielyGognyHFB+QRPADipoleStrength2018}. The fitted curves reproduce the experimental data well over the GDR energy range. In contrast, the microscopic QRPA predictions reproduce the gross trend. However, they systematically underestimate the peak magnitude near the GDR centroids and overestimate the strength on the high-energy tail. The peak parameters extracted from our fits are listed in \cref{tab:slo_params}.

\begin{table}[h]
  \caption{Summary of the fitted SLO $\gamma$SF parameters for $^{154,156,158,160}$Gd, grouped by peak index. The uncertainties represent one standard deviation.}
  \label{tab:slo_params}
  \text{(a) Peak 1 parameters:}
  \begin{ruledtabular}
    \begin{tabular}{lccc}
      Isotope & $\sigma_1$ (mb) & $E_1$ (MeV) & $\Gamma_1$ (MeV) \\
      \midrule
      $^{154}$Gd & 171.23$\pm$10.15 & 12.09$\pm$0.12 & 3.25$\pm$0.27 \\
      $^{156}$Gd & 204.74$\pm$23.77 & 12.24$\pm$0.18 & 3.06$\pm$0.46 \\
      $^{158}$Gd & 189.61$\pm$9.73 & 11.90$\pm$0.10 & 3.15$\pm$0.18 \\
      $^{160}$Gd & 213.10$\pm$6.77 & 12.21$\pm$0.04 & 2.71$\pm$0.13 \\
    \end{tabular}
  \end{ruledtabular}
  \vspace{0.15cm}
  \text{(b) Peak 2 parameters:}
  \begin{ruledtabular}
    \begin{tabular}{lccc}
      Isotope & $\sigma_2$ (mb) & $E_2$ (MeV) & $\Gamma_2$ (MeV) \\
      \midrule
      $^{154}$Gd & 241.56$\pm$13.97 & 15.12$\pm$0.09 & 2.81$\pm$0.15 \\
      $^{156}$Gd & 226.48$\pm$17.66 & 15.51$\pm$0.21 & 4.35$\pm$0.55 \\
      $^{158}$Gd & 245.38$\pm$12.75 & 15.13$\pm$0.08 & 2.72$\pm$0.13 \\
      $^{160}$Gd & 233.92$\pm$3.86 & 15.93$\pm$0.06 & 5.37$\pm$0.14 \\
    \end{tabular}
  \end{ruledtabular}
\end{table}

The fitted parameters show systematic trends across the Gd isotopic chain. The central energies $E_1$ and $E_2$ are relatively stable at approximately 12 MeV and 15 MeV, respectively. This two-peak structure is a characteristic feature of the GDR splitting in prolate deformed nuclei, where $E_1$ is attributed to oscillations along the nuclear symmetry axis and the higher-energy peak $E_2$ to oscillations perpendicular to it. The $\sigma_2$ is generally greater than $\sigma_1$ due to the difference in the degrees of freedom for these two oscillation modes. The widths, $\Gamma_1$ and $\Gamma_2$, ranging from 2.7 to 5.4 MeV, reflect the intrinsic damping of the collective dipole motion \cite{danosLongrangeCorrelationModel1958,bermanMeasurementsGiantDipole1975}.

\subsection{$M1$ $\gamma$-Ray Strength Function}
\label{sec:m1_rsf}

Although $E1$ transitions dominate the $\gamma$SF, the $M1$ component contributes may non-negligibly in the $2$--$10$ MeV region and affects neutron capture cross sections. We consider three phenomenological $M1$ models, and their impact on the calculated cross sections is discussed in Sec.~\ref{sec:cs}.

\noindent\textbf{(i) SLO Model.}
This prescription corresponds to the \texttt{strengthM1 1} option in TALYS. It adopts the RIPL-3 systematic formulae \cite{capoteRIPLReferenceInput2009} for the $M1$ resonance parameters:
\begin{align}
  f_{M1}(E_\gamma = 7~\text{MeV}) &= 1.58 \times 10^{-9} A^{0.47}, \nonumber\\
  E_{M1} &= 41\,A^{-1/3}~\text{MeV}, \qquad
  \Gamma_{M1} = 4~\text{MeV},
  \label{eq:m1_params}
\end{align}
where $A$ is the mass number, the $\sigma_{M1}$ is determined by inverting Eq.~\eqref{eq:psf} at $E_\gamma = 7$~MeV. This prescription serves as the default $M1$ input for TALYS calculations.

\noindent\textbf{(ii) Renormalization based on $E1$ strength.}
This prescription corresponds to the \texttt{strengthM1 2} option in TALYS. It utilizes the empirical coefficient $0.0588\,A^{0.878}$ given in the RIPL-2 handbook \cite{belgya2004reference} to normalize the $M1$ strength:
\begin{equation}
  f_{M1}(E_\gamma) = \frac{f_{E1}(E_\gamma)}{0.0588\,A^{0.878}},
  \label{eq:m1_ripl2}
\end{equation}

\noindent\textbf{(iii) Spin-flip and Scissors-mode.}

This prescription corresponds to the \texttt{strengthM1 3} option in TALYS. It models the total $M1$ strength as a superposition of two components. The Spin-flip component represents a broad resonance parameterized based on the nuclear mass. In contrast, the Scissors mode accounts for the extra strength observed in the $E_\gamma \sim 2$--$4$~MeV region. 
The scissors mode was first predicted by Lo Iudice and Palumbo \cite{iudiceNewIsovectorCollective1978} within a semiclassical two-rotor model. This collective excitation mode was subsequently observed experimentally in $^{156}$Gd by Bohle \textit{et al.} \cite{bohleNewMagneticDipole1984}
Since its strength is directly proportional to $\beta_2$, for the deformed Gd nuclei studied here ($\beta_2 \approx 0.3$--$0.35$) this low-energy contribution cannot be neglected.

\begin{figure}[h]
  \includegraphics[width=\columnwidth]{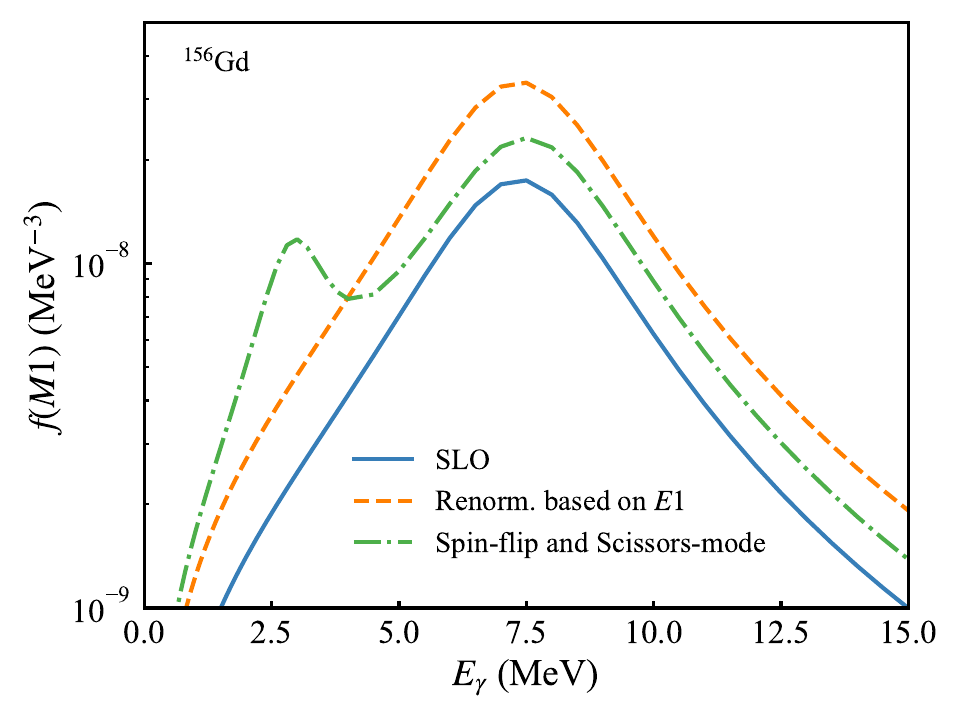}
  \caption{Comparison of the three $M1$ $\gamma$-ray strength function models for $^{156}$Gd. The plot shows the SLO model (blue solid line), the renormalization based on $E1$ strength function (orange dashed line), and the combined model with spin-flip and scissors-mode contributions (green dot-dashed line).}
  \label{fig:strengthM1}
\end{figure}

\cref{fig:strengthM1} compares these three $M1$ prescriptions for the nucleus $^{156}$Gd as an example. For the renormalization based on $E1$ strength (method \textbf{(ii)}), the calculation uses the best-fit $E1$ parameters from \cref{tab:slo_params}. All three models exhibit a resonance peak centered near $E_\gamma \approx 7.6$ MeV, with the method \textbf{(i)} predicting the lowest strength and the second model predicting the highest. The third model is distinguished by an additional pronounced enhancement in the low-energy region, $E_\gamma \sim 2$--$4$ MeV. These differences, particularly at low energies, illustrate the model uncertainty in the $M1$ strength function. This uncertainty propagates to the $(n,\gamma)$ cross section calculations.

\subsection{Nuclear Level Density Normalization}
\label{sec:nld}

At the neutron-separation energy $S_n$, the average s-wave neutron-resonance spacings $D_0$ \cite{capoteRIPLReferenceInput2009} are usually used to calculate the total level density $\rho(S_n)$. For the case of Gd, there are no experimental $D_0$ data for $^{160}$Gd, because $^{159}$Gd is unstable. To estimate reasonable $D_0$ for $^{160}$Gd we considered systematics of s-wave resonance spacings for this mass region by using the most recent evaluation of the RIPL-3. 
In addition, to determine $\rho(S_n)$, systematic errors arising from the spin distribution at $S_n$ must be reasonably accounted for\cite{larsenExperimentallyConstrained892016}.

Following Refs.~\cite{egidySystematicsNuclearLevel2005,vonegidyExperimentalEnergydependentNuclear2009}, we first describe the excitation-energy dependence of the spin-cutoff parameter $\sigma^2$ with two phenomenological Fermi-gas (FG) prescriptions. We list the rigid-body approximation of Ref.~\cite{egidySystematicsNuclearLevel2005} first, reads (FG05):
\begin{equation}
  \sigma_{\mathrm{FG05}}^{2}(E_x) = 0.0146\,A^{5/3}
  \frac{1 + \sqrt{1 + 4a\,(E_x - E_1)}}{2a},
  \label{eq:fg05}
\end{equation}
where $E_x$ is the excitation energy, $a$ is the NLD parameter, and $E_1$ is the excitation-energy back-shift determined from global systematics of Ref.~\cite{egidySystematicsNuclearLevel2005}. Second, we consider the FG spin-cutoff parameter of Ref. \cite{vonegidyExperimentalEnergydependentNuclear2009}(FG09):
\begin{equation}
  \sigma_{\mathrm{FG09}}^{2}(E_x) = 0.391\,A^{0.675}
  \left(E_x - \tfrac{1}{2}Pa^\ast\right)^{0.312},
  \label{eq:fg09}
\end{equation}
with $Pa^\ast$ denoting the pairing energy for deuteron as defined in Ref.~\cite{vonegidyExperimentalEnergydependentNuclear2009}. 
For the phenomenological $\sigma^2$, the spin distribution is given by the standard expression \cite{ERICSON1959481}:
\begin{equation}
  g(E_x,J) = \frac{2J + 1}{2\sigma^{2}(E_x)}
  \exp\!\left[-\frac{(J + \tfrac{1}{2})^{2}}{2\sigma^{2}(E_x)}\right],
  \label{eq:spin_dist}
\end{equation}
where $J$ is the spin of the levels at $E_x$. By using the phenomenological spin-cutoff parameters and assuming equal parity populations at $S_n$, the total level density is obtained from $D_0$ through
\begin{equation}
  \rho(S_n) =
  \frac{2\,\sigma^{2}(S_n)/D_0}
  {(J_\text{t} + 1)\exp\!\left[-\dfrac{(J_\text{t} + 1)^{2}}{2\sigma^{2}(S_n)}\right]
  + J_\text{t} \exp\!\left[-\dfrac{J_\text{t}^{2}}{2\sigma^{2}(S_n)}\right]},
  \label{eq:rho_sn}
\end{equation}
where $J_\text{t}$ is the ground-state spin of the target nucleus.

For the $^{154,156,158,160}$Gd nuclei, we investigated the spin distribution using $\sigma^2_{\mathrm{FG05}}$, $\sigma^2_{\mathrm{FG09}}$. We observe that the spin distribution calculated using $\sigma^2_{\mathrm{FG05}}$ is significantly broader and centered at higher spins compared to that using $\sigma^2_{\mathrm{FG09}}$. 
This finding is consistent with the results reported by Larsen \textit{et al.} \cite{larsenExperimentallyConstrained892016}. Therefore, we consider $\rho_{\text{FG05}}$ (calculated using $\sigma^2_{\text{FG05}}$) as the upper limit and $\rho_{\text{FG09}}$ (calculated using $\sigma^2_{\text{FG09}}$) as the lower limit. Consequently, we establish an uncertainty range for $\rho(S_n)$ using $\rho_{\text{FG05}}$ and $\rho_{\text{FG09}}$. Based on this, we adopt the HFB+c level densities, which are renormalized to reproduce the known discrete levels and the $\rho(S_n)$ derived from Eq.~\eqref{eq:rho_sn}.

\begin{figure}[h]
  \includegraphics[width=\columnwidth]{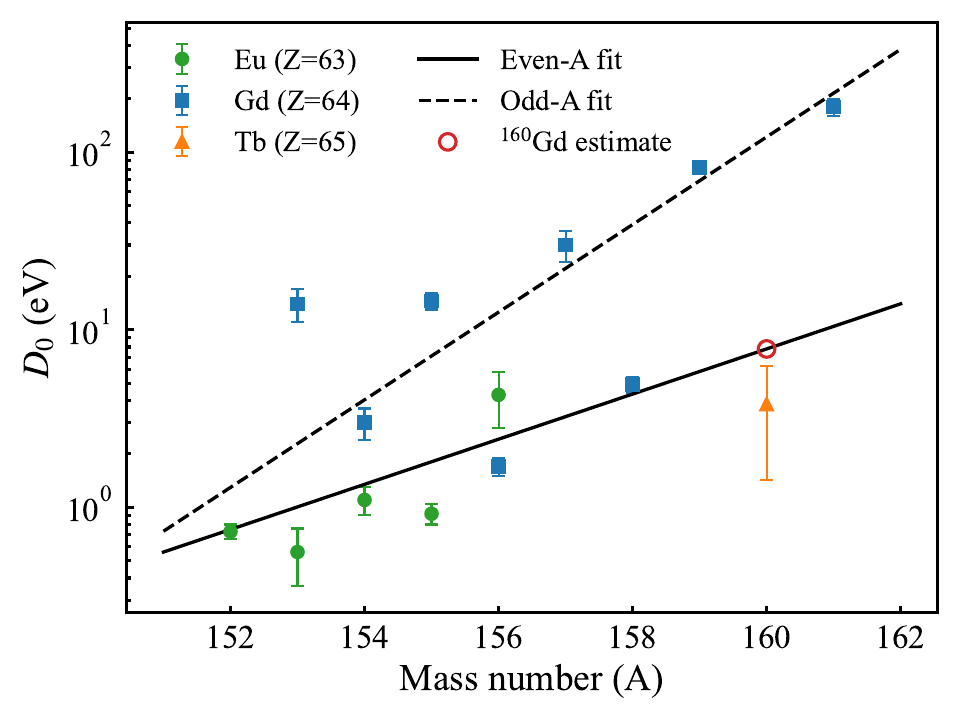}
  \caption{Systematics of $D_0$ for Gd isotopes and neighboring nuclei. The experimental data for Eu, Gd, and Tb are shown as green circles, blue squares, and orange triangles, respectively. The black solid line corresponds to the fit for even-$A$ nuclei, while the dashed line indicates the odd-$A$ fit. The red open circle marks represent the estimated $D_0$ for $^{160}$Gd from the even-$A$ fit.
  }
  \label{fig:D0_fit}
\end{figure}

As mentioned before,  the experimental $D_0$ data for $^{160}$Gd is unavailable. Given the limited experimental $D_0$ data for the Gd isotopic chain,
we estimate it by a linear fit of $\ln(D_0)$ versus $A$ using these neighboring nuclei (Eu, Gd, Tb) from RIPL-3. 
As shown in \cref{fig:D0_fit}, a strong odd-even effect is observed in the $D_0$ systematics. Therefore, separate fits were performed for odd-$A$ and even-$A$ nuclei. For $^{160}$Gd, the even-$A$ fit yields $D_0 = 7.8$~eV. Subsequently, with this estimated value and the experimental $D_0$ for the other isotopes, we calculated $\rho(S_n)$ for $^{154,156,158,160}$Gd using \cref{eq:rho_sn}, with spin cutoff parameters from \cref{eq:fg05,eq:fg09}.

We adopt the HFB+c level densities provided by Goriely \textit{et al.} \cite{gorielyImprovedMicroscopicNuclear2008} and renormalize them via a BO method as follows:

\begin{equation}
  \rho_{\text{norm}}(E_x,J,\pi)=\exp(c \sqrt{E_{ x } - \delta } ) \rho_{\text{HFB}}( E _ { x } - \delta, J, \pi ),
  \label{eq:HFB_nld}
\end{equation}
where $\delta$ is an energy shift, and constant $c$ acts as a slope correction, playing a role similar to that of the NLD parameter $a$ of phenomenological models.

\begin{figure}[h]
  \includegraphics[width=\columnwidth]{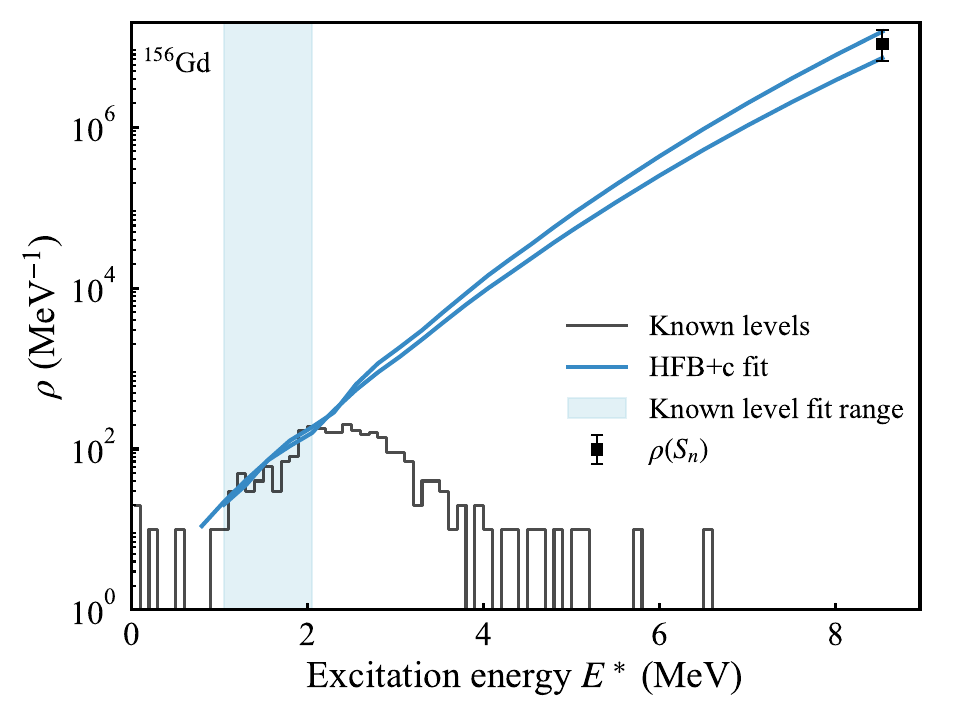}
  \caption{NLD for $^{156}$Gd, as an example. The known discrete levels are plotted as a black solid line. The upper and lower bounds of our renormalized HFB+c NLD are shown as two blue solid lines. The light blue shaded area indicates the energy range of the discrete levels used to fit. The $\rho(S_n)$ is shown as a black square with error bars.}
\label{fig:nld_fit}
\end{figure}

The optimal values of parameters $c$ and $\delta$ are determined through BO, with the objective of simultaneously satisfying two constraints. Specifically, at low excitation energies ($1.05 \le E_x \le 2.05$ MeV), $\rho_{\text{norm}}$ is fitted to level densities calculated from known levels using the binning method. The low-energy level density is defined as $\rho(E_{\text{center}}) = \Delta N / \Delta E$, where $\Delta N$ is the number of levels in a bin of width $\Delta E = 0.1$~MeV centered at $E_{\text{center}}$. Simultaneously, at $S_n$, $\rho_{\text{norm}}$ is fitted to the $\rho(S_n)$ calculated from \cref{eq:rho_sn}. To achieve this, we employ a Gaussian process surrogate model, minimizing the total $\chi^2$ of both constraints. The search ranges for parameters $c$ and $\delta$ are both set to $[-1.5, 1.5]$. The optimization is performed independently for the $\rho_{\text{FG05}}$ and $\rho_{\text{FG09}}$ of $\rho(S_n)$, yielding two $(c, \delta)$ parameter sets per isotope. These define the NLD uncertainty band. Final optimized parameters are listed in \cref{tab:nld_parameters}. As expected, $\rho_{\text{FG05}}$ yields more positive $\delta$ and larger $c$ than $\rho_{\text{FG09}}$, since a larger target value requires a steeper level density curve. \cref{fig:nld_fit} shows a fit for $^{156}$Gd as an example, where the renormalized curves reproduce the low-energy discrete levels while passing through the respective $\rho(S_n)$ anchors. Similar fit quality is obtained for the other isotopes. The uncertainty band bounded by these two curves is propagated to the cross section calculations. 

\begin{table}[h]
  \caption{Optimized parameters for the HFB+c level density renormalization, Eq.~\eqref{eq:HFB_nld}, for $^{154,156,158,160}$Gd. }
  \label{tab:nld_parameters}
  \begin{ruledtabular}
    \begin{tabular}{lcccc}
      \multirow{2}{*}{Isotope} & \multicolumn{2}{c}{$\rho_{\text{FG05}}$} & \multicolumn{2}{c}{$\rho_{\text{FG09}}$} \\
      \cmidrule(lr){2-3} \cmidrule(lr){4-5}
      & $c$ & $\delta$ (MeV) & $c$ & $\delta$ (MeV) \\
      \midrule
      $^{154}$Gd & 0.556 & -0.263& 0.330 & -0.740\\
      $^{156}$Gd & 0.781 &  0.015& 0.718 & -0.309\\
      $^{158}$Gd & 0.711 &  0.107& 0.378 & -0.356\\
      $^{160}$Gd & 0.596 &  0.131& 0.330 & -0.507\\
    \end{tabular}
  \end{ruledtabular}
\end{table}

\section{CROSS SECTION CALCULATIONS}
\label{sec:cs}

As discussed in the introduction, the NLD and $\gamma$SF are two key inputs for calculating neutron capture cross sections. We employ TALYS to calculate the $(n,\gamma)$ cross sections for the target nuclei $^{153,155,157,159}$Gd. Our approach consists of the following steps:

(1) Cross sections were calculated using all combinations of NLD and $\gamma$SF (both $E1$ and $M1$ components) models provided in TALYS. This procedure yields an estimate of the model uncertainty range, defined by the lower and upper limits from the theoretical models, for all isotopes studied here.

(2) Cross sections were calculated using the default models and parameters in TALYS.

(3) Cross sections and reaction rates were calculated by implementing the constrained $\gamma$SF and NLD parameters. The results for the stable isotopes $^{155,157}$Gd serve as a validation, supporting the reliability of the predictions for the unstable isotopes $^{153,159}$Gd.

\subsection{Cross sections predictions for model uncertainties}
\label{sec:cs_model_range}

\begin{figure*}[ht]
  \includegraphics[width=0.9\textwidth]{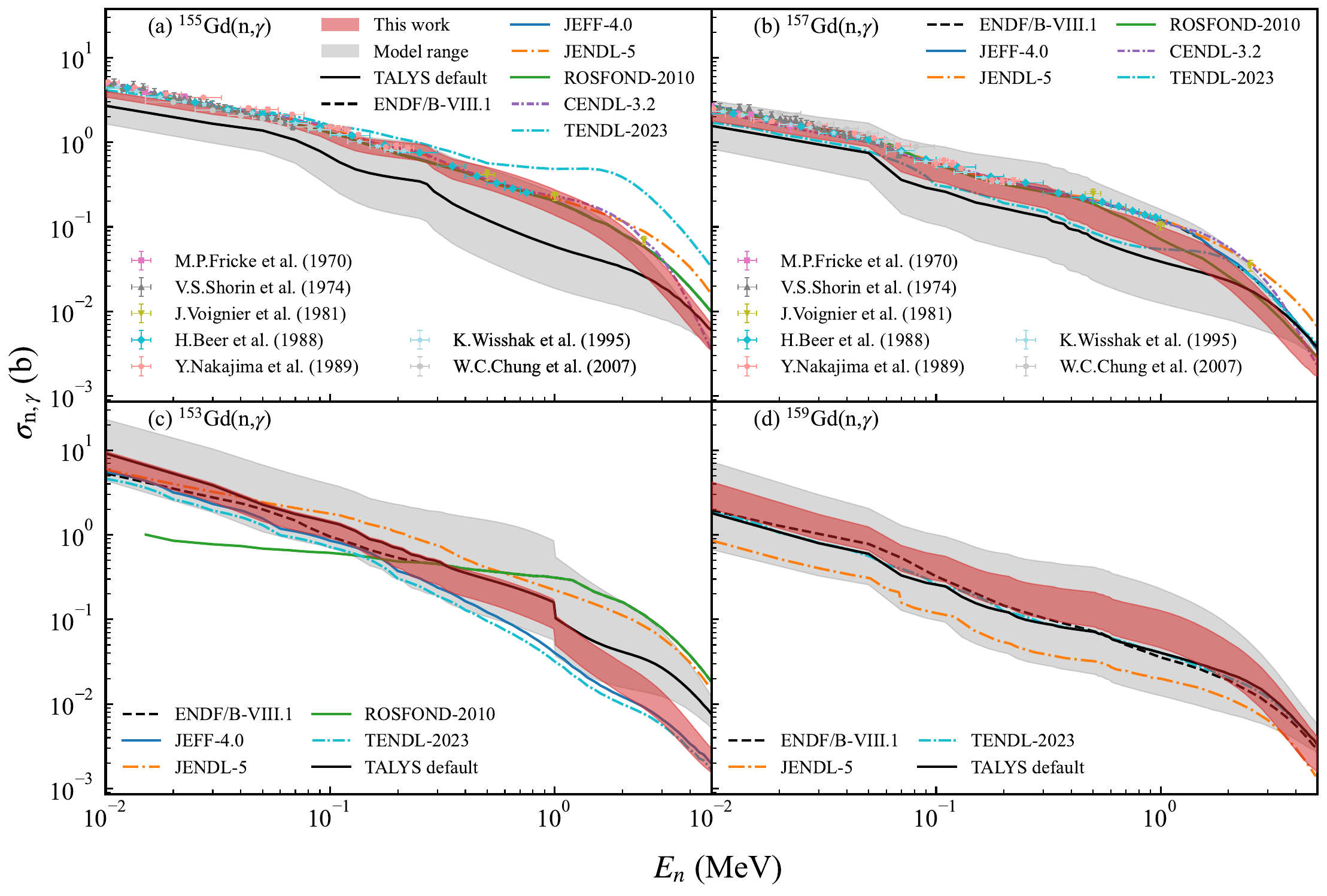}
  \caption{Panels (a)--(d) shows the $(n,\gamma)$ cross sections of $^{155,157,153,159}$Gd. The experimental datasets \cite{NAKAJIMA1989589,voignier1981neutron,shorin1974neutron,fricke1970measurements,wisshakStellarNeutronCapture1995,1988ApJ3311047B,정원정2007measurement} and main evaluation (ENDF/B-VIII.1, JEFF-4.0, JENDL-5, TENDL-2023, ROSFOND-2010, CENDL-3.2) \cite{nobre2024progress,plompenJointEvaluatedFission2020,iwamotoJapaneseEvaluatedNuclear2023,zabrodskaya2007rosfond,koningTENDLCompleteNuclear2019,ge2020cendl} are presented. The solid black lines represent the default prediction of TALYS. The red shaded bands represent the results of this work. The gray shaded bands show the model uncertainty range from all possible combinations of NLD, $E1$ $\gamma$SF, and $M1$ $\gamma$SF models available in TALYS.}
  \label{fig:xs_overview}
\end{figure*}

We performed calculations for all possible combinations of the available NLD models, $\gamma$SF prescriptions ($E1$ and $M1$) in TALYS. In these calculations, the Koning-Delaroche (KD) optical potential~\cite{koningLocalGlobalNucleon2003} was adopted. As the optical model potential is well-constrained by available experimental data, its contribution to the total uncertainty is negligible compared to that from the NLD and $\gamma$SF components. The resulting uncertainty range is shown as the gray shaded bands in \cref{fig:xs_overview}. The combinations yielding the minimum and maximum values are:

\textbf{(i) Minimum cross section:} calculated using the temperature-dependent RMF $E1$ strength function model from Daoutidis and Goriely \cite{PhysRevC.86.034328}, the SLO $M1$ strength function based on the RIPL-3 systematic formulae~\cite{capoteRIPLReferenceInput2009}, and the constant temperature + Fermi gas model for the NLD as introduced by Gilbert and Cameron \cite{doi:10.1139/p65-139} (TALYS keywords \texttt{strength 7}, \texttt{strengthM1 1}, \texttt{ldmodel 1}).

\textbf{(ii) Maximum cross section:} calculated using the Brink-Axel Lorentzian $E1$ strength function \cite{brinkIndividualParticleCollective1957,axelElectricDipoleGroundState1962}, the $M1$ strength function including spin-flip and scissors-mode contributions, and the NLD is based on temperature-dependent Hartree-Fock-Bogoliubov calculations using the Gogny force \cite{PhysRevC.86.064317} (TALYS keywords \texttt{strength 2}, \texttt{strengthM1 3}, \texttt{ldmodel 6}).

To quantify the individual contributions of model components to the total uncertainty, we varied the $\gamma$SF and NLD models separately while fixing other parameters. In the 0.01--5 MeV energy range, NLD variations yield an average uncertainty factor of $\sim$2.8 for the four Gd isotopes. The largest factors are obtained for $^{157,159}$Gd, reaching 3.4 and 3.2, respectively. The $E1$ $\gamma$SF produces an average uncertainty factor of $\sim$3.3, ranging from 2.7 for $^{153}$Gd to 4.4 for $^{159}$Gd. In contrast, the $M1$ strength function has a minimal impact, with a factor of only 1.1. This is attributed to the significantly smaller magnitude of the $M1$ strength compared to the dominant $E1$ component, thereby limiting its influence on the $(n,\gamma)$ cross sections.

Considered together, the range of predictions resulting from all possible combinations of NLD, $E1$ and $M1$ $\gamma$SF models corresponds to an average relative uncertainty (ARU) of approximately 167\% for the cross sections in this energy range. 
The ARU is defined as the relative uncertainty averaged over the considered energy range and the four isotopes. For a given isotope, it is calculated as 
$\textrm{ARU} = \frac{1}{N} \sum_{i=1}^{N} (\sigma_{\textrm{max},i}-\sigma_{\textrm{min},i})/ 2\langle\sigma_i\rangle$, where $N$ is the number of energy points, and $\sigma_{\textrm{max},i}$, $\sigma_{\textrm{min},i}$, and $\langle\sigma_i\rangle$ denote the maximum, minimum, and mean $(n,\gamma)$ cross sections at the $i$-th energy point.
These results underscore the importance of experimental constraints on both the $\gamma$SF and NLD to reduce the inherent model uncertainties in cross section predictions. This is particularly relevant for the energy region that dominates astrophysical $s$-process conditions and thermal reactor applications.

\subsection{TALYS default predictions}
For comparison, we also performed calculations using the default models and parameters of TALYS. These are the Simplified Modified Lorentzian model \cite{SMLO20081S} for $E1$ $\gamma$SF, the Spin-flip and scissors mode for $M1$ $\gamma$SF \cite{iudiceNewIsovectorCollective1978}, and the Constant Temperature + Fermi gas model \cite{gilbertCompositeNuclearlevelDensity1965} for the NLD. 
The results are displayed as black lines in \cref{fig:xs_overview}. For the stable isotopes $^{155,157}$Gd, significant discrepancies are observed between the TALYS default predictions and the experimental data. For $^{153}$Gd, the TALYS default prediction is higher than the major evaluated libraries when $E_n < 0.04$ MeV. In the high-energy region, the TALYS default falls in between the evaluated libraries. For $^{159}$Gd, the TALYS default prediction is in good agreement with the ENDF/B-VIII.1 and TENDL-2023 libraries.

\subsection{Results from constrained nuclear model parameters}

Finally, we employ our constrained nuclear model parameters to calculate the $^{153,155,157,159}\text{Gd}(n,\gamma)$ cross sections. For the neutron transmission coefficients, we adopted the KD optical potential~\cite{koningLocalGlobalNucleon2003}. The total uncertainty in our results accounts for three contributions: (1) $E1$ $\gamma$SF: Uncertainties are propagated via Monte Carlo sampling of the double-Lorentzian parameters (Table~\ref{tab:slo_params}) based on their covariance matrix. (2) $M1$ $\gamma$SF: The uncertainty is defined by the envelope of predictions from the three phenomenological models discussed in Sec.~\ref{sec:m1_rsf}.
(3) NLD: The uncertainty is propagated from the renormalized NLD bands derived in Section~\ref{sec:nld} (Table~\ref{tab:nld_parameters}).

We first calculate the cross sections for stable isotopes $^{155,157}$Gd$(n,\gamma)$, for which abundant experimental data exist, to validate our approach. As shown in Figs.~\ref{fig:xs_overview}(a) and (b), for the stable isotope $^{155}$Gd, our results show good agreement with both experimental data and most of the evaluated libraries across the entire energy range. The only exception is TENDL-2023, which begins to deviate from our results above 0.1~MeV. Similarly, for $^{157}$Gd, our predictions agree well with experimental data although our results are slightly lower than some data.
Based on this, we further calculate the $^{153,159}$Gd$(n,\gamma)$ cross sections, where no experimental data are available. As shown in Fig.~\ref{fig:xs_overview}(c) and (d), 
for $^{153}$Gd, our calculated cross sections are consistent with the ENDF/B-VIII.1 evaluation below 0.3~MeV. In the energy range of 0.3--1.2~MeV, our results lie between the JEFF-4 and JENDL-5 predictions. Above 1.2~MeV, our calculations agree well with TENDL-2023 and JEFF-4. 
Finally, for $^{159}$Gd, our calculations are slightly higher than most evaluations across the considered energy range. However, the lower bound of our calculated band shows good agreement with ENDF/B-VIII.1 and TENDL-2023. Overall, our results demonstrate that experimentally constraining the $\gamma$SF and NLD reduces the ARU to $\sim 30\%$. This represents a reduction by a factor of $\sim$5.5 compared to the uncertainty of the TALYS models.

\section{ASTROPHYSICAL IMPLICATION}
\label{sec:astro}

\begin{figure*}[ht]
  \includegraphics[width=0.98\textwidth]{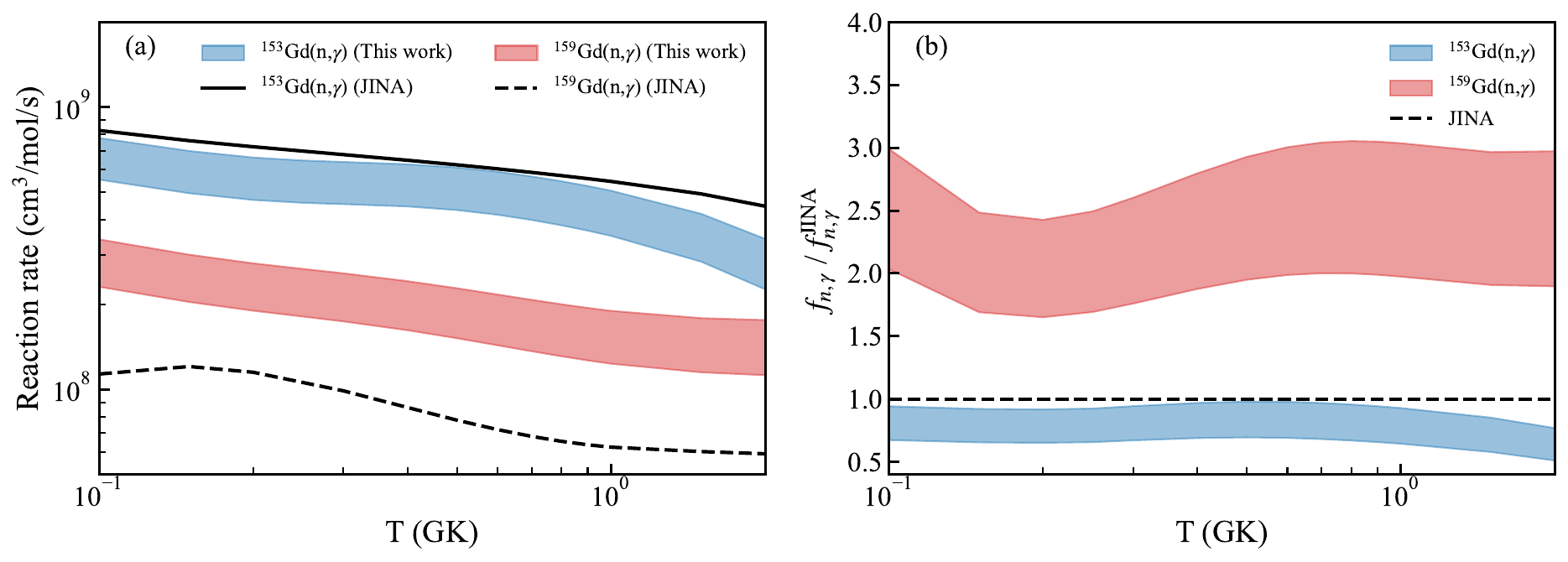}  
  \caption{Panels (a) and (b) show the astrophysical reaction rates and the $f_{n,\gamma}$ relative to JINA REACLIB, as a function of temperature, respectively. The blue shaded bands denote and the red shaded bands denote the $^{153,159}$Gd$(n,\gamma)$ result obtained in this work, respectively. In panel (a), the black solid and dashed lines show the JINA REACLIB recommendation for $^{153}$Gd and $^{159}$Gd, respectively. In panel (b), the horizontal dashed line at unity indicates agreement with JINA REACLIB.}

\label{fig:rate_brance}
\end{figure*}

Based on the constrained cross sections, we calculate the astrophysical reaction rates for $^{153,159}\text{Gd}(n,\gamma)$, as shown in Fig.~\ref{fig:rate_brance}(a).
These are compared with the recommended values from the JINA REACLIB database~\cite{cyburtJINAREACLIBDATABASE2010}.
Note that the \textit{ks03} evaluation~\cite{dillmann2009kadonis} is used for $^{153}$Gd and the \textit{thra} evaluation~\cite{RAUSCHER20001} is used for $^{159}$Gd in the database.
For $^{153}$Gd, the JINA recommended values agree with the upper limit of our uncertainty band.
In contrast, our calculated rates for $^{159}$Gd are notably higher than the JINA REACLIB recommended values, approximately by a factor of 2.9 at the mean value.
The enhanced $(n,\gamma)$ rate found in this work implies a stronger reaction flux through this channel than JINA REACLIB estimated, potentially facilitating the synthesis of heavier isotopes such as $^{160}$Gd.

\red{\subsection{The Branching at $^{153}$Gd and $^{159}$Gd}}

As mentioned in Sec.~\ref{sec:introduction}, the final abundances of the $s$-only Gd isotopes are sensitive to branchings at unstable nuclei, where neutron capture competes with $\beta$ decay. We investigate the branchings at $^{153}$Gd and $^{159}$Gd, which influence the production of the $s$-only isotope $^{154}$Gd and the $s$-process contribution to $^{160}$Gd, respectively. This competition can be quantified by the neutron capture branching ratio $f_{n,\gamma}$, defined as:

\begin{equation}
    f_{n,\gamma} = \frac{\lambda_{n,\gamma}}{\lambda_{n,\gamma} + \lambda_{\beta}} = \frac{n_n \langle \sigma v \rangle}{n_n \langle \sigma v \rangle + \ln 2 / t_{1/2}},
    \label{eq:branching}
\end{equation}
Here, $\lambda_{n,\gamma}$ and $\lambda_{\beta}$ denote the rates for neutron capture and $\beta$-decay, respectively. The $n_n$ is the neutron density, and $t_{1/2}$ is the $\beta$-decay half-life.
The $\langle \sigma v \rangle$ denotes the reaction rate per particle pair (in $\text{cm}^3 \text{s}^{-1}$).
Note that the astrophysical reaction rates presented in Fig.~\ref{fig:rate_brance}(a) are given as $N_A \langle \sigma v \rangle$ (in $\text{cm}^3 \text{mole}^{-1} \text{s}^{-1}$), and thus must be divided by Avogadro's constant $N_A$ for the evaluation of Eq.~(\ref{eq:branching}).

We calculated $f_{n,\gamma}$ over the temperature range of $0.1$--$2$ GK, with a typical $s$-process neutron density of $n_n = 10^8\,\text{cm}^{-3}$. 
The ratios $f_{n,\gamma}/f_{n,\gamma}^{\mathrm{JINA}}$ for $^{153}$Gd and $^{159}$Gd are shown in Fig.~\ref{fig:rate_brance}(b).
For $^{153}$Gd ($t_{1/2}=240.4$ d), we obtain $f_{n,\gamma}\approx 70\%$ at $T=0.2$ GK, meaning that neutron capture is the dominant channel, the ratio $f_{n,\gamma}/f_{n,\gamma}^{\mathrm{JINA}}$ remains close to unity. In contrast, for $^{159}$Gd ($t_{1/2}=18.5$ h), the short half-life leads to a much smaller branching ratio, $f_{n,\gamma}\approx 0.3\%$ at $T=0.2$ GK, i.e., $\beta$ decay dominates. Nevertheless, our constrained rates yield a larger branching ratio by a factor of about $2$--$3$ compared to JINA REACLIB. This notable enhancement implies a stronger neutron-capture flow toward $^{160}$Gd compared to the JINA REACLIB prediction.

\red{\subsection{Impact on $^{160}$Gd Production}}

The enhanced branching at $^{159}$Gd is expected to increase the $s$-process flow toward $^{160}$Gd. To quantify this impact, we perform $s$-process nucleosynthesis simulations as described below.
We performed $s$-process nucleosynthesis simulations using the single-zone reaction network code \texttt{NucNet Tools}~\cite{Meyer_nucnet_tools}. The calculations adopted fixed temperature, density, and neutron abundance. Specifically, the temperature was maintained at $T = 0.2$ GK, and the matter density was set to $\rho = 10^3$ g/cm$^3$. The neutron abundance was fixed at $Y_n = 1.66 \times 10^{-19}$, corresponding to approximately $n_n \approx 10^8$ cm$^{-3}$. The solar system elemental abundances from Ref.~\cite{lodders2003solar} were adopted as the initial composition. The reaction network was evolved for a duration of $3.15 \times 10^{12}$~s ($\approx 10^5$ yr), sufficient to reach equilibrium for the relevant isotopes.

\begin{figure}[h]
  \includegraphics[width=\columnwidth]{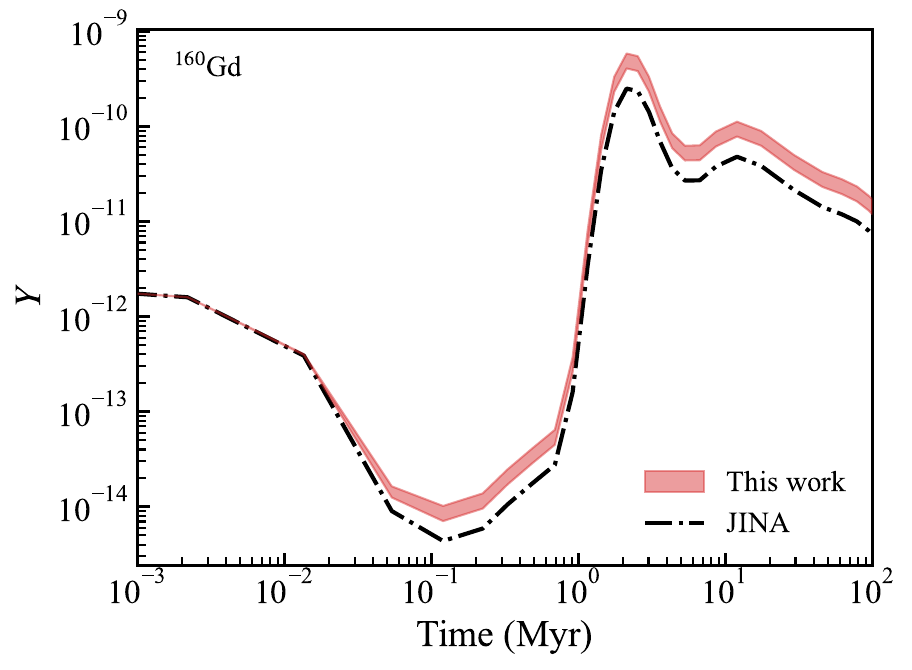}
  \caption{Evolution of $^{160}$Gd abundances in $s$-process nucleosynthesis. The red shaded band shows results using our updated $(n,\gamma)$ reaction rates, while the black dash-dotted line represents 
  the default prediction based on the JINA REACLIB rate.
  }
\label{fig:branching_abundance}
\end{figure}

As shown in Fig.~\ref{fig:branching_abundance}, the $^{160}$Gd abundance predicted in this work is consistently higher than the default prediction based on the JINA REACLIB rate after $2 \times 10^{-2}$ Myr.
In the intermediate and late stages of the evolution, our results exhibit a stable enhancement, stabilizing at a factor of approximately 2 compared to the default calculation. This result suggests that the $^{160}$Gd production in the $s$-process may have been underestimated in previous studies relying on the JINA REACLIB rates.
Therefore, an accurate $^{159}$Gd$(n,\gamma)$ rate is crucial for predicting Gd isotopic abundances in the $s$ process.
It should be noted that the present simplified single-zone model could be extended with more realistic stellar evolution conditions in future work, which would be valuable for further quantifying this impact.

\section{Summary and outlook}
\label{sec:summary}

In summary, we have successfully inferred the unstable $^{153,159}$Gd($n,\gamma$) cross sections by constraining the key $\gamma$SF and NLD inputs.
The resulting uncertainty is evaluated to be about $30\%$, which is smaller by a factor of 5.5 than the value of $\sim167\%$ predicted with different nuclear models in TALYS.
Note that this uncertainty constraints is particularly critical for investigating $s$-process branching points and the $i$-process, where experimental data are scarce, but accurate cross sections are essential.
Furthermore, we derived the astrophysical reaction rates for $^{153,159}$Gd isotopes. For the $^{153}$Gd$(n,\gamma)$ reaction, the reaction rates obtained are comparable to the JINA REACLIB recommendation.
However, the reaction rates for $^{159}$Gd$(n,\gamma)$ are found to be a factor of $\sim$2.9 higher than the JINA REACLIB recommendation. \red{As a result, the $s$-process branching ratio at $^{159}$Gd is enhanced relative to JINA REACLIB recommendation, leading to a factor of $\sim$2 higher $^{160}$Gd abundance in our $s$-process nucleosynthesis calculations.}

Our study enriches the ($n,\gamma$) data for the Gd isotopic chain, thereby providing support for the astrophysical reaction network calculations and nuclear science applications related to the Gd isotopes. 
In the future, we will apply our approach to derive $(n,\gamma)$ cross sections for other unstable isotopic chains. Sm and Er isotopes are of our priority, since their $(n,\gamma)$ data play an important role in determining cosmic-ray exposure conditions for planetary materials ~\cite{hidakaIsotopicVariationsSm2024}. Moreover, their accurate astrophysical reaction rates are indispensable for resolving the competition between $(n,\gamma)$ reactions and $\beta$-decays during the $r$-process freeze-out, which shapes the final rare-earth abundance pattern~\cite{kissMeasuringVdecayProperties2022}. We also should note that in order to infer the $(n,\gamma)$ data for these unstable isotopes of interest, we still need reliable experimental data of neighboring stable isotopes, which are useful for constraints of their $\gamma$SFs and NLDs. As a result, we call for more experimental measurements on $(n,\gamma)$ cross sections at operational and under-construction facilities including CSNS~\cite{chenChinasFirstPulsed2016,ZHOU2024169170}, n\_TOF~\cite{guerreroPerformanceNeutronTimeofflight2013}, LANSCE~\cite{HEIL2001229}, J-PARC~\cite{IGASHIRA201472}, GELINA~\cite{Willy2006GELINAAN}, and SARAF-II~\cite{KAIZER2022167001}.

\begin{acknowledgments}
This work was supported by the the National Key R\&D Program of China (Grant No: 2022YFA1603300), the National Natural Science Foundation of China (Grant Nos: 12575133 and 12505144).

\end{acknowledgments}

\bibliography{Gd_file}

\end{document}